\documentclass[english]{article}
\usepackage[T1]{fontenc}
\usepackage[latin1]{inputenc}
\usepackage{amssymb}

\makeatletter

\makeatletter

\makeatletter

\makeatletter

\makeatletter

\makeatletter

\makeatletter

\newtheorem{theorem}{Theorem}

\newtheorem{example}[theorem]{Example}

\newtheorem{remark}[theorem]{Remark}

\date{}

\makeatother

\makeatother

\makeatother

\makeatother

\makeatother

\makeatother

\usepackage{babel}
\makeatother
\begin{document}

\title{Reversibility and Non-reversibility in Stochastic Chemical Kinetics}

\author{V\.{} A\.{} Malyshev,\c{}S\.{} A\.{} Pirogov}

\maketitle
\begin{abstract}
Mathematical problems with mean field and local type interaction related
to stochastic chemical kinetics,\c{}are considered\.{} Our main
concern \={}various definitions of reversibility, their corollaries
(Boltzmann type equations, fluctuations,\c{}Onsager relations, etc\.{)}
and emergence of irreversibility. 
\end{abstract}
\tableofcontents{}

\section{Introduction}

It is known that any closed dynamical system has sufficiently simple
behavior -\={}it tends to equilibrium\.{} One of the reasons for
such a behavior is\={}time reversibility of the dynamics\.{} In
biology we have completely different,\c{}more complicated,\c{} behavior
Standard approach of biological and chemical theories to explain complicated
biological phenomena is to select some suitable non-reversible system
of chemical reactions. It is rather clear that in such a way any behavior
can be explained, that reduces the value of the theory. At the same
time, any non-reversible system of reactions can be formally deduced
from (reversible) laws of physics only under some scaling limits.

In this review we discuss mathematical problems related to the reversibility
and non-reversibility of chemical reaction systems.

In section 2 we consider connections between various probabilistic
characteristics of chemical reaction systems. We give classical and
stochastic description of chemical reaction system. Classical chemical
kinetics is the core of any theoretical and applied investigations
in biology. Stochastic chemical kinetics had always been more theoretical,
however also its applied aspect emerges, see \cite{ArRoAd,McAAr,GadLeeOth}.
We study how the fact of the invariance of Poisson measure is related
to 3 main variants if reversibility for chemical reaction systems:
chemical reversibility, reversibility as in Markov processes theory
and unitarity. At the end of the section we touch the (yet non developed)
case of large number of molecule types.

The main question is how chemical reaction systems behave for large
times. In section 3 it is shown that, under the reversibility assumption
this behavior is rather simple - the system tends to a fixed point,
that is to equilibrium. Dynamics in the vicinity of the fixed point
is normally described by two physical theories: linear perturbations
which bring us to Onsager relations, and even smaller perturbations
- stochastic fluctuations. We present here rigorous versions of these
theories. Finally, we present two ways of emergence of non-reversibility
i n chemical reaction systems - transport and time scaling.

In physics, a well-known procedure to get non-reversibility is the
limiting transition to Boltzmann equation. Stochastic and classical
chemical kinetics are typical examples of the theories which are called
mean field theories in physics. In section 4 we consider various possibilities
of how to to introduce locality in chemical reaction systems with
infinite number of particles. We specify 3 kinds of scaling which
give deterministic quasilinear partial differential equations with
nonlinear term similar to the right-hand part of the equations (\ref{kinetic_equ})
of classical chemical kinetics:

\begin{enumerate}
\item canonical scaling with $M\to\infty$ in chemical kinetics, where simultaneously
the particles move in the space; 
\item the Boltzmann-Grad limit in the particle system with local chemical
reactions; 
\item standard hydrodynamic scaling for particle systems performing random
walk on a lattice. 
\end{enumerate}
In Newtonian and quantum mechanics the notion of reversibility is
different and natural question is how this is related to previous
definitions. We touch this in section 5. Moreover, in all models of
chemical kinetics, considered in earlier sections, the reactions are
instant events. In fact however, reactions takes some time interval
which in both classical and quantum scattering theories is even formally
infinite. This means that scattering theory and chemical kinetics
lie on different timescales, sufficiently separated from each other.
In despite of this, the reaction intensities should be possible to
calculate from mathematical scattering theory. At the end of section
5 we introduce the notion of metastable particle, that is a particle
with finite life time, not as a singularity (resonance) in the spectrum
of a finite particle operator, but how some local formation in the
infinite particle system.

\section{Stochastic chemical kinetics}

\subsection{Stochastic and classical description of chemical reactions}

Assume there are finite molecule types $A_{v},v=1,...,V$. The chemical
reaction $r$ is defined by the vectors $d_{-}(r)=\{ d_{-}(v,r)\},d_{+}(r)=\{ d_{+}(v,r)\}$
of stoichiometric coefficients, where the non-negative integers $d_{-}(v,r)$
are the multiples- of the substrate $A_{v}$ of type $v$, $d_{+}(v,r)$
are the multiples- of the product of type $v$. The reactions is written
formally as\[
\sum_{v}d_{-}(v,r)A_{v}\to\sum_{v}d_{+}(v,r)A_{v}\]
 Consider the system of chemical reactions, that is the Markov process
$\mathcal{M}$ with the states $n=(n_{1},,...,n_{V})$ and $R$ transitions
(reaction types) $r=1,2,...,R$\begin{equation}
n_{v}\to n_{v}+d_{+}(v,r)-d_{-}(v,r),v=1,...,V\label{reaction}\end{equation}
 where $n_{v}$ is the number of molecules of type $v$.

Here we should do the following comment. Below in section 5, especially
in 5.3, we discuss how the given stochastic model of chemical kinetics
is related to the basic principles of physics. Note however that there
is no rigorous deduction of this model from basic postulates. Here
two problems appear. The first one is related in stochastic chemical
kinetics the time scale is coarser comparative with the time of one
reaction. This time can be calculated using quantum scattering theory.
However, even for the simplest reactions such calculations are sufficiently
complicated and, moreover, the separation of these two scales had
never been realized in a rigorous way. This would be sufficient to
deduce the local models considered below in sections 4, from the first
principles. However in sections 2 and 3 we consider mean field models.
And the second problems is how the mean field models are related to
local models. We do not give answer to this question, which appears
always when in physics one uses mean field models. One should say
however that the limiting Boltzmann equations in mean field and local
case differ only by renormalization of the reaction intensities. An
intermediate case could be the Kac model where the molecule can react
not with each other, but only with molecules situated on a distance
not bigger than $\epsilon L$, where $L$ is the diameter of the system.

The reaction $r$, that is the transition (\ref{reaction}), has the
rate (speed of the reaction in chemical kinetics)\begin{equation}
\lambda_{r}(n)=\lambda_{r}(n\to n-d_{-}(r)+d_{+}(r))=M^{-m_{-}(r)+1}a_{r}\prod_{v\in I(r)}n_{v}...(n_{v}-d_{-}(v,r)+1),\label{intensity}\end{equation}
 where $a_{r}$ are the fixed numbers (constants of the speed in chemical
kinetics), $M$ is the scaling coefficient, and\[
m_{-}(r)=\sum_{v}d_{-}(v,r),I(r)=\{ v:d_{-}(v,r)>0\}\]
 The power of $M$ in (\ref{intensity}) corresponds to the so called
canonical scaling, as defined in \cite{Ser1}.

Note that equation (\ref{intensity}) expresses the law of mass action,
well-known in chemical kinetics.

We will also consider distinguished reactions {}``input'' and {}``output''\[
\emptyset\to A_{v},A_{v}\to\emptyset\]
 with the rates $Ma_{v,in}$ and $a_{v,out}n_{v}$ correspondingly.
If there are no such reactions in the system, we will call the system
closed, otherwise open.

Note shortly that the history of stochastic chemical kinetics starts
with the paper \cite{Leo}, the subsequent story consisted mainly
in the study of systems wit small $R$ and $V$, see references in
\cite{McQua,Kal}.

\subsubsection{Convergence to the equations of classical kinetics}

Classical kinetics can be obtained from the stochastic one in the
limit $M\to\infty$. Namely, we consider the family of processes $n_{v}(t)=n_{v}^{(M)}(t)$,
depending on the parameter $M$. We cite here the well-known result,
see references in \cite{Ser1}, under more general assumptions, including
possible input and output.

\begin{theorem}

Assume that initially the limits \[
c_{v}(0)=\lim_{M\to\infty}\frac{n_{v}(0)}{M}\]
 exist for all $v$. Then for any $v$ and $t>0$ there exist the
following limits in probability\[
c_{v}(t)=\lim_{M\to\infty}\frac{n_{v}(t)}{M}\]
 which satisfy the following equations (the standard equations of
chemical kinetics)

\begin{equation}
\frac{dc_{v}(t)}{dt}=\sum_{r:v\in O(r)}d_{+}(v,r)a_{r}\prod_{w\in I(r)}c_{w}^{d_{-}(w,r)}-\sum_{r:v\in I(r)}d_{-}(v,r)a_{r}\prod_{w\in I(r)}c_{w}^{d_{-}(w,r)}\label{kinetic_equ}\end{equation}
 where\[
O(r)=\{ v:d_{+}(v,r)>0\}\]
 \end{theorem}

The conservation laws ensure compactness. Let $w_{bv}$ be the number
of atoms of type $b$ in the molecule of type $v$. Then for all reactions
$r$ (except input and output) and all atom types $b$ one has\[
\sum_{v\in I(r)}d_{-}(v,r)w_{bv}=\sum_{v\in O(r)}d_{+}(v,r)w_{bv}\]
 Note that the limiting terms for input and output are\[
\sum_{v}(a_{v,in}-a_{v,out}c_{v})\]

If there are no conservation laws then in general the trajectory can
go to infinity or can reach a fixed point for finite time. The proof
of the theorem for this case see in \cite{Ser1}.

\subsection{Unitarity and invariance of the Poisson measure }

The condition for the invariance of the Poisson measure for the Markov
process $\mathcal{M}$\[
\mu(n)=\prod_{v}\frac{\overline{b}_{v}^{n_{v}}}{n_{v}!}e^{-\overline{b}_{v}},\overline{b}_{v}=Mb_{v}\]
 where $b_{v}$ are some fixed parameters, can be written as the equality
of output\begin{equation}
F_{out}=\mu(n)\sum_{r}\lambda_{r}(n)\label{out}\end{equation}
 and input\begin{equation}
F_{in}=\sum_{r}\mu(n')\lambda_{r}(n'),n'=n+d_{-}(r)-d_{+}(r)\label{in}\end{equation}
 probability flows for any given state $n=(n_{1},,,,.n_{V})$. Note
that by (\ref{intensity}) the summation in the right-hand side of
(\ref{out}) is only over admissible reactions that is on the reaction
such that $n_{v}-d_{-}(v,r)\geq0$, and in the right-hand side of
(\ref{in}) is over $r$ such that $n'_{v}-d_{-}(v,r)=n{}_{v}-d_{+}(v,r)\geq0$.

Using\[
\frac{\mu(n')}{\mu(n)}=\prod_{v}\overline{b}_{v}^{n'_{v}-n_{v}}\frac{n_{v}!}{n'_{v}!},\]
 where $F_{in}=F_{out}$, we get, dividing by $\mu(n)$ and using
the uniqueness of defining $n_{v}'$ in terms of $n_{v}$ and $r$:
$n'_{v}-d_{-}(v,r)=n_{v}-d_{+}(v,r)$\[
\sum_{r}M^{-m_{-}(r)+1}a_{r}\prod_{v}\overline{b}_{v}^{n'_{v}-n_{v}}\frac{n_{v}!}{n'_{v}!}n'_{v}...(n'_{v}-d_{-}(v,r)+1)=\]
 \[
=\sum_{r}M^{-m_{-}(r)+1}a_{r}\prod_{v\in I(r)}n_{v}...(n_{v}-d_{-}(v,r)+1)\]
 or\[
\sum_{r}M^{-m_{-}(r)+1}a_{r}\prod_{v\in I(r)}\overline{b}_{v}^{d_{-}(v,r)-d_{+}(v,r)}n_{v}...(n_{v}-d_{+}(v,r)+1)\]
 \[
=\sum_{r}M^{-m_{-}(r)+1}a_{r}\prod_{v\in I(r)}n_{v}...(n_{v}-d_{-}(v,r)+1)\]
 Finally\[
\sum_{r}M^{-m_{+}(r)+1}a_{r}\prod_{v\in I(r)}b_{v}^{d_{-}(v,r)-d_{+}(v,r)}n_{v}...(n_{v}-d_{+}(v,r)+1)\]
 \begin{equation}
=\sum_{r}M^{-m_{-}(r)+1}a_{r}\prod_{v\in I(r)}n_{v}...(n_{v}-d_{-}(v,r)+1)\label{uslovie}\end{equation}
 where $m_{+}(r)=m_{-}(r)-\sum_{v\in I(r)}(d_{-}(v,r)-d_{+}(v,r))$.

\begin{theorem}

The following statements are equivalent:

1) The Poisson measure $\mu$ is invariant under the given system
of chemical reactions;

2) (Stueckelberg condition, or unitarity condition) for any vector
$d$ one has\begin{equation}
\sum_{r:d_{-}(r)=d}a_{r}\prod_{v\in I(r)}b_{v}^{d_{-}(v,r)}=\sum_{r:d_{+}(r)=d}a_{r}\prod_{v\in I(r)}b_{v}^{d_{-}(v,r)}\label{Unitarity}\end{equation}

\end{theorem}

Proof. By reversibility of the previous calculation we have proved
that condition 1) is equivalent to the condition (\ref{uslovie}).
Let us use the fact that for the polynomial of $n_{v}$ equal zero
it is necessary and sufficient that all coefficients of its quasimonoms
were zero. Two equal monoms are defined by the vector\[
d=d_{-}(r)=d_{+}(r)\]
 Note also that the scaling coefficients cancel, as $m_{-}(r)=m_{+}(r)$.

Note that condition (\ref{Unitarity}) has sense also for classical
deterministic chemical kinetics where it can be interpreted as follows:
the sum of reaction rates creating the (and only this) group of particles
characterized by the vector $d$, is equal to the sum of reaction
rates which annihilate this (and only this) group.

A. N. Rybko remarked that, under the condition of the existence of
invariant Poisson measure, time reversal in a process of stochastic
chemical kinetics brings to the process of the same type, but in general
with other set of chemical reactions.

\subsection{Reversibility in probability theory}

Probability theory has its own notion of time reversibility. Stationary
random process $X_{t}$ is called time reversible, if the finite-dimensional
distributions of the processes $X_{t}$ and $Y_{t}=X_{-t}$ coincide.
For Markov processes (for example, with denumerable number of states
and continuous time) this definition is equivalent to the following
detailed balance condition\begin{equation}
\pi_{i}\lambda_{_{ij}}=\pi_{j}\lambda_{ji}\label{detailed_prob}\end{equation}
 where $\lambda_{ij}$ are the transition rates, $\pi$ is the invariant
measure of the Markov process. The latter definition can be applied
also to the time homogeneous but not stationary, that is to null recurrent
and non-recurrent Markov processes, but only if there exists (now
infinite) non-negative measure $\pi$ on the state space, satisfying
condition (\ref{detailed_prob}). This will be called reversibility
with respect to measure $\pi$. If $\pi$ equals one in each point,
then reversibility is equivalent to the symmetry of the matrix $(\lambda_{ij})$.
That is why any reversible chain can be obtained from the chain with
symmetric transition matrix by the transformation \[
\lambda_{ij}\to w_{i}\lambda_{ij}w_{j}^{-1},w_{i}=\sqrt{\pi_{i}}\]

Kolmogorov criterion of reversibility of the Markov process with respect
to measure $\pi$ consists in the fulfillment of the inequalities
\[
\lambda_{i_{1}i_{2}}\lambda_{i_{2}i_{3}}...\lambda_{i_{n}i_{1}}=\lambda_{i_{1}i_{n}}\lambda_{i_{n}i_{n-1}}...\lambda_{i_{2}i_{1}}\]
 for any sequence of states $i_{1},...,i_{n}$. Under this condition,
using definition (\ref{detailed_prob}), measure $\pi$ is easily
constructed as\[
\pi_{i_{n}}=\pi_{0}a_{0i_{1}}a_{i_{1}i_{2}}...a_{i_{n-1}i_{n}},a_{ij}=\frac{\lambda_{ij}}{\lambda_{ji}}\]
 for any sequence of states $i_{1},...,i_{n}$, where\[
\pi_{0}^{-1}=1+\sum_{k}a_{G(0,k)}\]
 where the sum is over all states $k\neq0$, $G(0,k)$ is some path
from $0$ to $k$, that is a sequence of states $i_{0},i_{1},...,i_{n}$
such that $i_{0}=0,i_{n}=k$, and\[
a_{G(0,k)}=a_{0i_{1}}a_{i_{1}i_{2}}...a_{i_{n-1}i_{n}}\]

\subsection{Chemical reversibility}

The notion of chemical reversibility can be introduced in the general
framework of Markov chains. Assume that on a given state space $X$
be given some finite set $\mathbf{A}$ of transition matrices $\lambda_{mn}^{\alpha},m,n\in X,\alpha\in\mathbf{A}$
for some continuous time Markov chains. The elements of the set $\mathbf{A}$
will be called {}``chemical reactions''. Assume that on $\mathbf{A}$
the involution operation is defined - the reverse reaction $\alpha\to\alpha'\neq\alpha$,
so that $(\alpha')'=\alpha$.

Define the Markov chain $\xi_{\mathbf{A}}$ on $X$ by the transition
rates\[
\lambda_{mn}=\sum_{\alpha}\lambda_{mn}^{\alpha}\]
 We call such chain chemically reversible, if there exists a probability
distribution $\pi_{n}$ on $X$, that is for all $\alpha$ the following
condition is holds\[
\pi_{n}\lambda_{nm}^{\alpha}=\pi_{m}\lambda_{mn}^{\alpha'}\]

In classical chemical kinetics the inverse reaction $r'$ to the reaction
$r$ is uniquely defined by the conditions $d_{-}(r)=d_{+}(r'),d_{-}(r')=d_{+}(r)$.

For classical kinetics the detailed balance condition - that is the
equality of the rates of direct and inverse reactions - looks as follows\begin{equation}
a_{r}\prod_{v\in I(r)}b_{v}^{d_{-}(v,r)}=a_{r'}\prod_{v\in I(r')}b_{v}^{d_{-}(v,r')}\label{detailed_chem}\end{equation}
 We shall see the difference between conditions (\ref{detailed_prob})
and (\ref{detailed_chem}).

\begin{theorem}Let the chemical reaction system be given where for
any reaction $r$ there exists an inverse reaction $\overline{r}$,
Then the following conditions are equivalent:

1) The corresponding Markov process is chemically reversible;

2) The chemical reaction system is reversible (as the random process)
with respect to some Poisson measure.

\end{theorem}

Proof.

($1\to2$) For a pair $r,r'$ of two mutually inverse reactions the
condition of chemical reversibility is \begin{equation}
\pi_{n}a_{r}\prod_{v}\frac{n_{v}!}{(n_{v}-d_{-}(v,r))!}=\pi_{n'}a_{r'}\prod_{v}\frac{n'_{v}!}{(n'_{v}-d_{+}(v,r))!}\label{chemrev}\end{equation}
 for some measure $\pi_{n}$ and for any $n,n'=n-d_{-}(r)+d_{+}(r)$.
Denote \[
z_{n}=\log(\pi_{n}\prod_{v}n_{v}!)\]
 then instead of (\ref{chemrev}) we have\begin{equation}
z_{n}-z_{n'}=\log\frac{a_{r'}}{a_{r}}\label{equil}\end{equation}
 Denote $l_{r}$ the right-hand side of (\ref{equil}), it is called
the equilibrium constant in chemistry. Is we assume that in the given
chemical reaction system $\mathbf{A}$ all vectors\[
d=d(r)=d_{-}(r)-d_{+}(r),r\in\mathbf{A}\]
 are different then we can consider $l_{r}$ as a function $l(d)$
of $d=d(r)$. However it can occur that different reactions have equal
vectors $d(r)$. Anyway, we have the following properties of $l(d)$:

1. from the right-hand side of (\ref{equil}) and from $d(r)=-d(r')$,
it follows $l(d)=-l(-d)$;

2. from the left-hand side of (\ref{equil}) it follows: if for some
sequence of reactions $r_{1},...,r_{k}$ one has $\sum_{i=1}^{k}d(r_{i})=0$,
then\[
\sum_{i=1}^{k}l(d(r_{i}))=0\]
 From these two properties it follows that $l(d)$ can be extended
to an (additive) homeomorphism $\phi$ of additive subgroup $Q\subset Z^{V}$
to $R$, where $V$ is the number of molecule types in the system,
and $Q$ is generated by all vectors $d(r)$, that is \[
\phi(0)=0,\phi(x+y)=\phi(x)+\phi(y)\]
 This homomorphism can be extended to the additive homomorphism of
$Z^{V}$ to $R$. Any such homomorphism can be written as\[
\phi(x)=\sum_{v}\alpha_{v}x_{v}\]
 for some real numbers $\alpha_{v}$.

Now one can check that the Poisson measure\begin{equation}
p_{n}=\prod_{v}\frac{\alpha_{v}^{n_{v}}}{n_{v}!}e^{-\alpha_{v}}\label{poisson}\end{equation}
 satisfies condition (\ref{chemrev}). We have\[
w_{n}=\log(p_{n}\prod_{v}n_{v}!)=-\sum_{v}\alpha_{v}n_{v}+const\]
 and hence, for any pair of vectors $n,n'$ such that $n'=n-d_{-}(r)+d_{+}(r)$,
we get\[
w_{n}-w_{n'}=\sum_{v}\alpha_{v}d_{v}(r)=l(d(r)),l(d(r))=\log\frac{a_{r'}}{a_{r}}\]
 It follows that the measure $p_{n}$ satisfies equation (\ref{equil})
or, that is equivalent (\ref{chemrev}). Hence, the considered Markov
chain has this measure as a stationary distribution, moreover it is
reversible with respect to this measure.

($2\to1$) Reversibility condition of the Markov process with respect
to the Poisson measure (\ref{poisson}) is that for any two vectors
$n,n'$\begin{equation}
p_{n}\sum_{r}a_{r}\prod_{v}\frac{n_{v}!}{(n_{v}-d_{-}(r))!}=p_{n'}\sum_{r'}a_{r'}\frac{n'_{v}!}{(n'_{v}-d_{+}(r))!}\label{poisson_obr}\end{equation}
 where the summation is over all reactions $r$ such that $n_{v}-n'_{v}=d_{-}(r)-d_{+}(r)$
and $n_{v}\geq d_{-}(r)$ for all $v$. From (\ref{poisson_obr})
it follows that\[
\prod_{v}\alpha_{v}^{n_{v}}\sum_{r}a_{r}\prod_{v}\frac{1}{(n_{v}-d_{-}(r))!}=\prod_{v}\alpha_{v}^{n'_{v}}\sum_{r}a_{r'}\prod_{v}\frac{1}{(n'_{v}-d_{+}(r))!}\]
 Hence\[
\sum_{r}a_{r}\prod_{v}\alpha_{v}^{d_{-}(r)-d_{+}(r)}\frac{1}{(n_{v}-d_{-}(r))!}=\sum_{r}a_{r'}\prod_{v}\frac{1}{(n_{v}-d_{-}(r))!}\]
 under the same agreement concerning summation. Otherwise speaking\[
\sum_{r}a_{r}\prod_{v}\alpha_{v}^{d_{-}(r)-d_{+}(r)}n_{v}...(n_{v}-d_{-}(r)+1)=\]
 \[
=\sum_{r}a_{r}\prod_{v}n_{v}...(n_{v}-d_{-}(r)+1)\]
 for fixed $n_{v}$ and $d(r)$. As $n_{v}$ are arbitrary, the latter
equation can hold only if for the vector $d_{-}(r)$ (and corresponding
vector $d_{+}(r)=d(r)+d_{-}(r)$) the following equality holds\[
a_{r}\prod_{v}\alpha_{v}^{d_{-}(r)-d_{+}(r)}=a_{r'}\]
 This condition is obviously equivalent to the chemical reversibility
for the Poisson measure (\ref{chemrev}) and to the detailed balance
condition (\ref{detailed_chem}) as well.

\begin{remark}

The Schloegl example, see below section 3.1, shows, that the stochastic
reversibility does not imply the invariance of a Poisson measure.

\end{remark}

\subsection{Large number of types}

We do not give here exact formulations but indicate interesting classes
of problems. The models where $V$ is large, physically may correspond
to two quite different situations, that we consider below. In the
first case the system is subdivided into marge number of cells. One
can imagine that the cells form some space subdivision. In each cell
the number of substance (molecule) types is bounded. Any substance
can react only with the substances from the same cell or neighbor
cells, the number of neighbor cells is also bounded. In the second
case the number of substance types in each cell may be large. Moreover,
even if the number of atom types is bounded, the number of atoms in
a molecule can be very large. Even more, most interesting situations
appear when the reaction rates are large, that is the life time of
these substances are short. Then all substances influence the evolution
of the system. Substances with short life time correspond to the so
called metastable molecules or clusters, as they are called in chemistry.

\paragraph{Locally finite reaction networks}

We will introduce them for the case when\[
\sum_{v}d_{+}(v,r),\sum_{v}d_{-}(v,r)\]
 are uniformly bounded in $r$ , and\[
\sum_{r}d_{+}(v,r),\sum_{r}d_{-}(v,r)\]
 are uniformly bounded in $v$. Then $V$ can be taken countable.

The infinite system of equations of classical kinetics can be deduced
as above. Of course, $M$ again has the sense of the volume but not
the whole volume, but only local volume of each cell.

Unitarity condition and theorem 2 can be generalized to such infinite
case. Markov reversibility condition is well-known for the locally
interacting process, our system may fit to this case. In the same
spirit the chemical reversibility condition can be generalized.

\paragraph{Models with large clusters }

Let we have only one type of elementary atoms and the clusters differ
only by size. Formally speaking, we consider the partitions of the
set with $N$ elements-atoms into subsets which we call clusters.
Let $m_{v}$ be the number of clusters of size $v=1,...,V$, that
is consisting of $v$ atoms, we assume the conservation law\[
\sum_{v}vm_{v}=N\]
 but the number $N$ initially can be random, the number of types
$V$ is fixed, but finally we will be interested with the asymptotics
when $V\to\infty$. Define the Markov chain, the states of which are
the vectors $m=(m_{1},...,m_{N})$. Possible reactions: 1) $r_{n}$
- appending of a particle to a cluster of size $n>1$\[
m_{1}\to m_{1}-1,m_{n}\to m_{n}-1,m_{n+1}\to m_{n+1}+1,\]
 and joining $r_{1}$ of the two particles\[
m_{1}\to m_{1}-2,m_{2}\to m_{2}+1\]

2) $r_{n}'$ - separation of a particle from the cluster of size $n>2$.\[
m_{1}\to m_{1}+1,m_{n}\to m_{n}-1,m_{n-1}\to m_{n-1}+1\]
 and decay $r_{1}'$ of two-particle cluster \[
m_{1}\to m_{1}+2,m_{2}\to m_{2}-1\]
 Their rates $a_{r_{n}}=a_{n},a_{r_{n}'}=a_{n}'$ do not depend of
course on $V,N$. Assume that that chemical reversibility condition
holds for any pair $r_{n},r_{n+1}'$. Denote $\pi$ the stationary
Poisson distribution for this chain with parameters $b_{v}$. Then
the chemical reversibility condition gives for all $n=1,2,...,V$\[
a_{n}b_{1}b_{n}=a_{n+1}'b_{n+1}\]
 These relations allow to construct many examples. In particular,
for any given sequence of positive integers $b_{1},b_{2},...$ one
can uniquely find the fractions $\frac{a_{n}}{a_{n}'}$ of direct
and inverse reaction rates. Some examples of this kind with concrete
$a_{n},a_{n}'$ one can find in the book \cite{Kelly}, where however
small ensembles are considered.

\section{Complexity of behavior and non-reversibility}

\subsection{Fixed points}

Consider first the equations (\ref{kinetic_equ}) of classical chemical
kinetics. It appears that under the unitarity condition (hence, under
reversibility conditions) their solutions have sufficiently simple
behavior.

The Boltzmann entropy of the positive vector of concentrations $c=(c_{1},...,c_{V})$
with respect to some fixed non-negative measure $c^{0}$ on $\{1,...,V\}$
is defined as\[
H(c)=H(c,c^{0})=\sum_{v}c_{v}\ln\frac{c_{v}^{0}}{c_{v}}+\sum_{v}c_{v}=\sum_{v}c_{v}\ln\frac{ec_{v}^{0}}{c_{v}}\]
 Note that the factor $e$ cannot be omitted, if the number of particles
is not conserved. It is interesting that this expression is quite
similar to the expression of entropy for mixture of ideal gases, see
\cite{LanLif,Ser3}.

\begin{theorem}

Assume that for a given chemical reaction system there exists vector
$c^{0}>0$, with respect to which the system satisfies the unitarity
condition (\ref{Unitarity}). The the following assertions hold:

1) for any solution $c_{v}(t)$ of kinetic equations (\ref{kinetic_equ})
with any initial condition $c_{v}(0)$ the entropy is non-decreasing,
that is\[
\frac{dH(c(t))}{dt}\geq0\]

2) as $t\to\infty$ any solution $c(t)$ of equations (\ref{kinetic_equ})
tends to some fixed point $c_{\infty},$ which in general depends
on $c(0)$;

3) unitarity condition holds for any fixed point of equations (\ref{kinetic_equ});

4) if for some fixed point the detailed balance condition holds, it
also holds for any other fixed point.

\end{theorem}

Proof of these assertions can be found in \cite{Ser1}.

\paragraph{Systems with one particle type}

Without unitarity condition such simple behavior is rarely possible.
It is instructive to consider the following examples with one particle
type. For one particle type and reactions $r:d_{-}(r)\to d_{+}(r)$
we have\begin{equation}
\frac{dc(t)}{dt}=\sum_{r}(d_{+}(r)-d_{-}(r))a_{r}c^{d_{-}(r)}(t)\label{schlogl}\end{equation}
 where in the right-side part there can be arbitrary polynomial of
$c$. We are interested only in non-negative solutions of this equation.
Consider a particular case, the Schloegl model \cite{Volk}, where
there are only reactions of types $0\to1,1\to0,2\to3,3\to2$. Denote
the corresponding $a_{r}$ by $a_{01},a_{10},a_{23},a_{32}$. It is
not difficult to show that the unitarity condition holds only in the
following three cases:

\begin{enumerate}
\item {}``input-output'': $a_{23}=a_{32}=0$; 
\item $a_{01}=a_{10}=0$ - we call this case the closed Schloegl model; 
\item if\[
\frac{a_{23}}{a_{32}}=\frac{a_{01}}{a_{10}}\]

\end{enumerate}
One can show also that in these cases not only unitarity condition
holds but also the detailed balance condition (\ref{detailed_chem}).

When in general first four coefficients of the polynomial \ref{schlogl}
are different from zero, then on the positive half-axis there can
be either one or three fixed points. In the latter case the right
and left fixed points are stable, and the middle fixed point is unstable.

\begin{remark}

For several molecule types the class of polynomials, corresponding
to reactions with or without conservation laws, is sufficiently wide.
As an example consider the set $R(I)$ of all reactions $r$, where
$I=I(r)=O(r)$. Then\[
\sum_{r\in R(I)}a_{r}(d_{+}(v,r)-d_{-}(v,r))\prod_{v\in I}c_{v}^{d_{-}(v,r)}\]
 that is we have an arbitrary polynomial.

It follows that non-reversible chemical reaction systems may have
behavior of arbitrary complexity. For example, many systems with cycles
are known \cite{Murray}. It seems one can prove that chemical reaction
systems can model any algorithmic behavior (that is any finite automata).
From one side, it confirms unbounded possibilities of biological systems,
and from other side it can be a source of artificial adjustment of
the theory to experiment.

\end{remark}

\subsection{Fluctuations and Onsager relations}

The behavior of systems in a neighborhood of a fixed point (that is
close to the equilibrium) was much studied in statistical physics.
Thus in the fluctuation theory for equilibrium dynamics (that is dynamics
conserving the equilibrium measure) one considers the neighborhoods
of the concentrations $c_{v}$ of the order $\frac{1}{\sqrt{M}}$,
one can also perform linearization of the dynamical system in a small
neighborhood - this is one of the ways to get Onsaget relations. Now
we give exact definitions.

\paragraph{Linearization and Onsager relations}

Consider small perturbations of the initial data $n_{v}(0)$, that
is assume that the limits\[
c_{v}(0)=\lim_{M\to\infty}\frac{n_{v}^{(M)}(0)}{M}\]
 belong to a small $\epsilon$-neighborhood of the fixed point.

The linearization of the equations (\ref{kinetic_equ}) around some
distinguished solution $c_{v}(t)$ gives the following equations for
the variations $x_{v}(t)$ of the solution $c_{v}(t)$\[
\frac{dx_{v}}{dt}=\sum_{r}a_{r}(d_{+}(v,r)-d_{-}(v,r))\sum_{u}d_{-}(u,r)c_{u}^{d_{-}(u,r)-1}x_{u}\prod_{w\neq u}c_{w}^{d_{-}(w,r)}\]
 In particular, if as a distinguished solution we take a fixed point
$c_{v}(t)=\overline{c}_{v}$, then we have the system of linear differential
equations with constant coefficients\begin{equation}
\frac{dx_{v}}{dt}=\sum_{r}a_{r}(d_{+}(v,r)-d_{-}(v,r))\sum_{u}d_{-}(u,r)\overline{c}_{u}^{d_{-}(u,r)-1}x_{u}\prod_{w\neq u}\overline{c}_{w}^{d_{-}(w,r)}\label{eqVAR}\end{equation}
 Rewrite it as\begin{equation}
\frac{dx_{v}}{dt}==\sum_{u}\lambda_{vu}x_{u}\label{linearkinetic}\end{equation}
 where the matrix of coefficients is\begin{equation}
\lambda_{vu}=\sum_{r}a_{r}(d_{+}(v,r)-d_{-}(v,r))d_{-}(u,r)\overline{c}_{u}^{d_{-}(u,r)-1}\prod_{w\neq u}\overline{c}_{w}^{d_{-}(w,r)}\label{eqVAR0}\end{equation}
 \begin{theorem}

Under the chemical reversibility condition the following Onsager relations
hold\begin{equation}
\lambda_{vu}\overline{c}_{u}=\lambda_{uv}\overline{c}_{v}\label{Onsag}\end{equation}

\end{theorem}Proof. From (\ref{eqVAR0}) we have\begin{equation}
\lambda_{vu}\overline{c}_{u}=\sum_{r}a_{r}(d_{+}(v,r)-d_{-}(v,r))d_{-}(u,r)\prod_{w}\overline{c}_{w}^{d_{-}(w,r)}\label{eqVAR1}\end{equation}
 Assume now that the detailed balance condition holds both for direct
and inverse reactions $r,r'$\begin{equation}
a_{r}\prod_{w}\overline{c}_{w}^{d_{-}(w,r)}=a_{r'}\prod_{w}\overline{c}_{w}^{d_{+}(w,r)}\label{DB1}\end{equation}
 Then the contribution to the expression (\ref{eqVAR1}) from two
mutually inverse reactions $r,r'$ equals\[
a_{r}(d_{+}(v,r)-d_{-}(v,r))d_{-}(u,r)\prod_{w}\overline{c}_{w}^{d_{-}(w,r)}+\]
 \begin{equation}
+a_{r'}(d_{-}(v,r)-d_{+}(v,r))d_{+}(u,r)\prod_{w}\overline{c}_{w}^{d_{+}(w,r)}\label{eqVAR2}\end{equation}
 The contribution from the same reactions to the expressions for $\lambda_{uv}\overline{c}_{v}$,
similar to (\ref{eqVAR1}), equals\[
a_{r}(d_{+}(u,r)-d_{-}(u,r))d_{-}(v,r)\prod_{w}\overline{c}_{w}^{d_{-}(w,r)}+\]
 \begin{equation}
+a_{r'}(d_{-}(u,r)-d_{+}(u,r))d_{+}(v,r)\prod_{w}\overline{c}_{w}^{d_{+}(w,r)}\label{eqVAR3}\end{equation}
 It is easy to see that the expressions (\ref{eqVAR2}) and (\ref{eqVAR3})
coincide. In fact, the terms containing the product's $d_{-}(u,r)d_{-}(v,r)$
and $d_{+}(u,r)d_{+}(v,r)$, in (\ref{eqVAR2}) and (\ref{eqVAR3})
are identical, and the remaining terms are equal correspondingly to
\[
a_{r}d_{+}(v,r)d_{-}(u,r)\prod_{w}\overline{c}_{w}^{d_{-}(w,r)}+a_{r'}d_{-}(v,r)d_{+}(u,r)\prod_{w}\overline{c}_{w}^{d_{+}(w,r)}\]
 and\[
a_{r}d_{+}(u,r)d_{-}(v,r)\prod_{w}\overline{c}_{w}^{d_{-}(w,r)}+a_{r'}(d_{-}(u,r)d_{+}(v,r)\prod_{w}\overline{c}_{w}^{d_{+}(w,r)}\]
 But due to detailed balance condition (\ref{DB1}) these expressions
also coincide.

\paragraph{Entropy}

If the unitarity condition of Stueckelberg holds, then one explain
the increase of the entropy $H(c)$ in a sufficiently simple and intuitive
way. In fact, in this case the Poisson measure $Q$ with the parameters
$M\overline{c}_{i}$ is invariant. One can indicate a model where
the dynamics leaves invariant the class of all Poisson measures, exactly
(see section 4.1) pr approximately. Thus, the Poisson measure $P_{t}$
can change with time, and its parameters $Mc_{i}(t)$ depend on time
somehow. Then the Kullback-Leibler divergence equals to \[
\rho_{KL}(Q,P_{t})=\int P_{t}\ln\frac{P_{t}}{Q}=-M\widetilde{H}(c)\]
 where\[
\widetilde{H}(c)=H(c)-\sum\overline{c}_{i}\]
 (it is evident that $H(\overline{c})=\sum_{i}\overline{c}_{i}$).
On the other hand it is well known, that for any Markov process, under
some technical conditions, $\rho_{KL}(Q,P_{t})$ is monotone time
decreasing, although possibly not strictly monotone. In this sense
theorem 5 is a detalization of this general assertion.

The entropy $\widetilde{H}(c)$ can be also interpreted in terms of
probabilities of large deviations (where $n$ is an arbitrary configuration)\[
\lim_{M\to\infty}\frac{\ln Q(n)}{M}=\widetilde{H}(c)\]
 if $\frac{n}{M}\to c$.

\paragraph{Fluctuations}

The fluctuations of the concentrations at time $t$, for fixed $M$,
are defined as \begin{equation}
\xi_{v}^{(M)}(t)=\xi_{v}^{(M.n_{v}(0))}(t)=\frac{n_{v}(t)-<n_{v}(t)>}{\sqrt{M}}\label{fluct}\end{equation}
 This process depends, moreover, on the initial data $n_{v}(0)$,
or from $<n_{v}(t)>$. Equilibrium fluctuations correspond to the
process $\xi_{v}(t+s)$ in the limit $s\to\infty$, when there is
a stable fixed point $c_{v}^{fixed}=\overline{c}_{v}$ and\begin{equation}
\lim_{M\to\infty}\frac{n_{v}(0)}{M}=c_{v}^{fixed}\label{uniquefixed}\end{equation}

\begin{theorem}Let the Markov process $\xi_{v}^{(M)}(t)$, for fixed
$M$, be stationary and let the condition (\ref{uniquefixed}) hold.
Then the limit (in the sense of finite-dimensional distributions)
$(\xi_{v}(t))$ of the vector process $(\xi_{v}^{(M)}(t))$ as $M\to\infty$
is the Ornstein-Uhlenbeck process with mean $0$ and covariance matrix\[
D_{vv'}(t-t'))=<\xi_{v}(t)\xi_{v'}(t')>\]
 defined by the formula (\ref{covariance}) below.

\end{theorem}

Proof. For any fixed time, the vector $(\xi_{v}^{(M)}(t))$ has Poisson
distribution, thus it converges in probability to a Gaussian vector
$(\xi_{v}(t))$ which has moreover independent components and\[
<\xi_{v}(t)>=0,<\xi_{v}^{2}(t)>=\overline{c}_{i}\]
 Otherwise speaking the probability density of the vector $(\xi_{v}(t))$
is\[
\rho=const\exp(-\frac{1}{2}\sum_{i}\frac{x_{i}^{2}}{\overline{c}_{i}})\]
 The quadratic form in the exponent coincides with the quadratic part
of the Taylor expansion of the function $H(\overline{c})$ at the
point $c=\overline{c}$. That is, the covariance matrix $\beta^{-1}$
is diagonal with diagonal elements $\overline{c}_{i}$. By definition,
the matrix of kinetic coefficients $\gamma=\lambda\beta^{-1}$. Thus
it is\[
\gamma_{ij}=\lambda_{ij}\overline{c}_{j}\]
 By theorem 7 this matrix is symmetric as expected.

Now we will describe the time correlations. If the limit of the process
$\xi_{v}^{(M)}(t)$ exists, then it is a stationary and reversible
process, as it is stationary and reversible for all $M$. From central
limit theorem it follows even more, see (\cite{EthKurtz,BKPR}). Namely,
in the limit $M\to\infty$ this process is Gaussian with the generator\[
L=\sum_{v,w}D_{vw}\frac{\partial^{2}}{\partial x_{v}\partial x_{w}}+\sum_{v}b_{v}\frac{\partial}{\partial x_{v}}\]
 where\[
b_{v}(x)=\sum_{u}\lambda_{vu}x_{u}\]
 that is the drift is defined by linearized kinetic equations (\ref{linearkinetic}),
see also \cite{Spohn2}.

This can be checked by a straightforward calculation: the drift at
point $(n_{1},...,n_{V})$ is equal to\[
S_{v}(n_{1},...,n_{V})=\]
 \begin{equation}
=\sum_{r}(d_{+}(v,r)-d_{-}(v,r))a_{r}M^{-m(r)+1}\prod_{w}n_{w}(n_{w}-1)...(n_{w}-d_{-}(w,r)+1)\label{drift}\end{equation}
 Put\[
c_{v}^{(M)}=\frac{n_{v}}{M},s_{v}=\frac{S_{v}}{M}\]
 Consider the stable fixed point $\overline{c}_{v}^{(M)}=\frac{\overline{n}_{v}}{M}$,
where in particular \[
s_{v}=O(\frac{1}{M})\]
 In the $M^{-\frac{1}{2}}$-neighborhood of this point, more exactly
for the points $q_{v}M^{-\frac{1}{2}}$, let us find the vector field
of drifts, already up to $M^{-\frac{1}{2}}$. For this substitute\[
c_{v}=\overline{c}_{v}+x_{v}M^{-\frac{1}{2}}\]
 into (\ref{drift}). The resulting expression is similar to the expression
for linearized system (\ref{eqVAR}). The conservation laws holds
here as well. Hence, in the limit we get a Markov process with linear
drift, that is Ornstein-Uhlenbech process, as the unique reversible
stationary Gaussian process.

The diffusion matrix can be found from the condition of stationarity
of the Gaussian measure\[
L^{*}\rho=0\]
 This condition leads to the equation\[
D\beta=-\lambda\]
 or\[
D=-\lambda\beta^{-1}=-\gamma\]
 Otherwise speaking, the diffusion matrix coincides with the matrix
of kinetic coefficients. This connection between $D$ and $\beta$
holds of course in the general theory of equilibrium fluctuation as
well \cite{LanLif}. The multi-time covariance\[
\phi_{vw}(t)=<\xi_{v}(t)\xi_{w}(0)>,t>0\]
 can be found from the system of linear equations\[
\frac{d\phi}{dt}=\lambda\phi\]
 with initial condition $\phi(0)=\beta^{-1}$. Thus\begin{equation}
\phi(t)=e^{\lambda t}\beta^{-1}\label{covariance}\end{equation}
 .The symmetry of the matrix $\phi$ is obvious. In fact, Onsager
relations can be written as $\lambda'=\beta\lambda\beta^{-1}$, where
the prime means transposition. Thus\[
e^{\lambda't}=\beta e^{\lambda t}\beta^{-1}\]
 which means that the matrix $e^{\lambda t}\beta^{-1}$ is symmetric.

\paragraph{Kubo formula}

Consider the matrix of kinetic coefficients $\gamma_{uv}=\lambda_{uv}\overline{c}_{v}$.
In a neiborhood of the fixed point let us compare the quadratic covariance
form and the quadratic part of the Boltzmann entropy. Both depend
on the chemical reaction system and are defined by the reduction procedure
of the quadratic forms with the conservation laws.

For the speed process\[
J_{v}(t)=\frac{d}{dt}\xi_{v}(t)\]
 its covariance matrix\[
\vartheta_{vw}(t)=<J_{v}(t)J_{w}(0)>\]
 is\[
\vartheta(t)=-\frac{d^{2}}{dt^{2}}\phi(t)=-\lambda^{2}\phi(t),t>0\]
 From this the famous Kubo formula (more exactly its classical variant)
follows\[
\int_{0}^{\infty}\vartheta(t)dt=\gamma\]
 where $\gamma$ is the matrix of kinetic coefficients.

\subsection{How non-reversible reaction systems come into play}

The problem of time non-reversibility is the central problem in physics,
and even more in biology. Thus it would important to understand what
elementary sources of non-reversibility could be in general.

Non-reversibility is related to the openness of the system, in particular
with transitions like transport. For example, introduction of inputs,
outputs or transport channels between reactions, also introduction
of complementary substances. This can be realized physically by external
fields (gravity, electromagnetic fields), current, in which the reactions
occur, or with difference of diffusion constants.

Other source of non-reversibility are various analogs of the famous
Boltzmann idea, called the Boltzmann-Grad limit in mathematical physics.
There are many other parameters in closed systems the scaling of which
allows to get non-reversible systems. For example, large deviations
in the system can be considered as a special scaling in the initial
state, when the system is driven far away from equilibrium and then
one can follow the path of reaching the equilibrium.

\subsubsection{Transport}

Let we have several closed system of chemical reactions. One could
join them together, allowing transport of substance from one system
$i$ to another $j$. We will say then that there is a transport channel
from $i$ to $j$. We want to show that after appending transport
to the system it becomes in general non-reversible.

Consider a closed chemical reaction system, satisfying the detailed
balance condition (\ref{detailed_chem}) with parameters $b_{v}$
of the Poisson measure. Introduce in addition inputs and outputs,
as mutually inverse reactions, for some substances with parameters
$a_{in,v},a_{out,v}$. For the system to stay chemically reversible
these parameters should satisfy the following restriction \[
b_{v}=\frac{a_{in,v}}{a_{out,v}}\]
 Consider now two independent closed systems and connect it by the
transport channel with rates\[
a_{12,v}n_{1,v},a_{21,v}n_{2,v}\]
 where $n_{i,v}$ is the number of molecules of type $v$ in the system
$i=1,2$. Again we get that the condition (\ref{detailed_chem}) will
hold only for one value of the parameters $\frac{a_{12,v}}{a_{21,v}}$.

\subsubsection{Constant concentrations}

In chemical kinetics one often encounters the assumption that one
or more substances have constant concentrations. The first question
is when the chemical reaction system with constant concentrations
of some substances is non-reversible.

Consider a reversible system $R$ of chemical reactions with substance
from the set $V$. Let $b_{v}$ be the concentrations, satisfying
the detailed balance condition. Fix somehow the concentrations $c_{v}$
from the set $W\subset V$, in general they different from $b_{v}$.
Consider the reduced system of reactions with the substances from
the set $V\setminus W$, assuming that there is no reactions where
enter only substances from $W$.

Then the numbers $\phi(v)=\frac{c_{v}}{b_{v}}$ should satisfy the
equation\[
\prod_{v}\phi(v)^{d_{-}(v,r)}=\prod_{v}\phi(v)^{d_{+}(v,r)}\]
 for any reaction $r\in R$. Thus, the function $h(v)=\ln\phi(v)$
is an additive first integral for any reaction $r\in R$, that is\[
\sum_{v}d_{-}(v,r)h(v)=\sum_{v}d_{+}(v,r)h(v)\]
 Here $h(v),v\in W,$ are some given positive numbers, the rest $h(v),v\notin W,$
are variables. Denote $\mathcal{L}$ the set of all additive integrals
of our system of chemical reactions, and $\mathcal{H}_{W}$ is the
subspace, consisting of the functions $f(v)$ such that $f(v)=\ln\frac{c_{v}}{b_{v}}$
for $v\in W$. The dimension of this subspace equals $|V|-|W|$. Thus,
if $|W|>\dim\mathcal{L}$, then generically (that is for some open
everywhere dense set of values $c_{v},v\in W$) the intersection $\mathcal{L}\cap\mathcal{H}_{W}$
is empty. The exceptional values of the parameters $c_{v},v\in W,$
are the projection of the subspace $\mathcal{L}$ on the coordinate
plane $\{ c_{v},v\in W\}$ and cannot fill in all this plane if $|W|<\dim\mathcal{L}$.
Thus, if in the reversible system $R$ we fix $k+1$, where $k=\dim\mathcal{L},$
then the resulting system in general will not be reversible.

\begin{example}

For a system of reactions with atoms $C,O,H$ of carbon, oxygen and
hydrogen one has three first integrals - conservation of the numbers
of these atoms. For sufficiently reach systems of reactions in organic
chemistry there are no other independent integrals. Thus here $k+1=4$.

\end{example}

\begin{remark}

If there are reactions with substances only from $W$, then they can
be excluded from the list of reactions of the reduced system. If the
exclusion of these reactions does not change $\dim\mathcal{L}$ (for
example, if the reaction $r$ can be changed to a chain of reaction),
then the previous conclusion persists, as they are based only on the
comparison of $\dim\mathcal{L}$ and $|W|$.

Above it was essential that, up to the exclusion of these reactions,
the detailed balance condition of the reduced system coincides with
the detailed balance condition for the initial system. For the unitarity
condition and the fixed points this is already not true.

\end{remark}

Formally, one can get constant concentration, for example of one substance
$v_{0}$, using various scaling limits. One of the possibilities is
to take the concentration $v_{0}$ big enough, moreover in the rates
(\ref{intensity}) of the Markov process one substitutes $Mn_{v_{0}}$
instead of $n_{v_{0}}$, or, that is the same, to subtract $d_{-}(v_{0},r)$
from $m_{-}(r)$. Then the limiting kinetic equations \[
\frac{dc_{v}(t)}{dt}=f(c_{1},...,c_{V})\]
 will be\[
\frac{dc_{v}(t)}{dt}=f(c_{v},v\neq v_{0},c_{v_{0}})\]
 \[
\frac{dc_{v_{0}}(t)}{dt}=0,c_{v_{9}}(0)=c_{v_{0}}\]
 Alternative way is to introduce input and output $v_{0}$, which
have large rates $a_{v,in},a_{v,out}$, so that the stationary distribution
of the process {}``input-output'' has a fixed value. Evidently,
we will get constant concentration in the limit.

\paragraph{Inverse problem}

How to get given non-reversible reaction from reversible ones ? Consider
some non-reversible system $R_{n}$ consisting of several pairs of
mutually inverse reactions. They are defined by the rates $a_{r}$.
Note that for some values $a_{r}$ this reaction system is reversible.
Assume that in each reaction $r$ some substances (enzymes) $w=w(r)$
participate. That is why in the expression (\ref{intensity}) for
the reaction rates there should be factor $c_{w}$. If one assumes
as above that $c_{w(r)}$ are constant, then the rates $a_{r}$ change
to $a_{r}c_{w(r)}$, and by adjusting the concentrations $c_{w(r)}$
one can get necessary values $a_{r}$.

\section{Stochastic local models}

\subsection{Stochastic models with mixed dynamics}

Classical chemical kinetics is the typical example of what is called
mean field theory in physics. Also, mixed models are possible where
the reactions are described by mean field models, but some local transport
is introduced as well. These models are useful for joining together
chemical kinetics and chemical thermodynamics in ONE microscopic model.
In the models introduced in \cite{Ser3}, the mean field dynamics
for chemical reactions is complemented by the free movement of particles.
These models allow to deduce main laws of chemical thermodynamics.
Here we describe only the limiting dynamics itself, referring the
reader for all applications to \cite{Ser3,Ser4}.

\paragraph{Dynamics of finite system of particles}

Firstly, define the dynamics of finite particle system in the volume
$\Lambda\subset R^{3}$ with periodic boundary conditions. At time
$t=0$ we throw $N$ particles to this volume, uniformly and independently,
where $N$ is random and has Poisson distribution with the density
$\frac{<N>}{\Lambda}=c$. To each particle we prescribe independently
the type $j$ and velocity $v$, which has density $p_{0}(j,v)$ (with
respect to Lebesgue measure)\[
\sum_{j}\int p_{0}(j,v)dv=1\]
 Assume for simplicity that only binary reactions exist\begin{equation}
r=(j,v),(j_{1},v_{1})\to(j',v'),(j_{1}',v_{1}')\label{bina}\end{equation}
 If the velocities took only finite number of values then the chemical
reaction system could be defined exactly as in section 1. In case
of continuous velocities instead of parameters $a_{r}$ in (\ref{intensity})
we introduce the rate densities, that is integrable (in all four variables)
functions\[
a_{r}(((j_{1},v_{1}),(j_{1}',v_{1}')|(j,v),(j',v'))\geq0\]
 Thus, each particle $i$ changes its type and velocity at random
moments\[
t_{i1}<t_{i2}<...\]
 In the intervals between these moments the particle moves with constant
velocity, obtained in the latter reaction.

\paragraph{Dynamics of the infinite system}

Consider the set $\mathbf{X}$ of countable locally finite configurations
$X=\{ x_{i},v_{i},j_{i}\}$ of particles in $R^{3}$, where each particle
$i$ has coordinate $x_{i}$, velocity $v_{i}$ and type $j_{i}$.
Denote $\mathfrak{M}$ the set of all probability measures on $\mathbf{X}$
with the following properties:

\begin{itemize}
\item the coordinates of these particles are distributed as the homogeneous
Poisson point field on $R^{3}$ with some fixed density $c$; 
\item the vectors $(j_{i},v_{i})$ are distributed with common density $p(j,v)$,\[
\sum_{j}\int_{R^{3}}p(j,v)dv=1,\]
 independently of the coordinates and other particles. 
\end{itemize}
Random dynamics on $\mathbf{X}$ is given by the pair $(\mathbf{X}^{0,\infty},\mu)$,
where $\mu=\mu^{0,\infty}$ is the probability measure on the set
$\mathbf{X}^{0,\infty}$ of countable arrays $X^{0,\infty}(t)=\{ x_{i}(t),v_{i}(t),j_{i}(t)\}$
of piecewise linear trajectories $x_{i}(t),v_{i}(t),j_{i}(t)),0\leq t<\infty$.
It is assumed that the measure $\mu$ belongs to the family $\mathfrak{M}^{0,\infty}$
of measures on $X^{0,\infty}(t)$, defined by the following properties:

\begin{itemize}
\item if for any $t$ we denote $\mu(t)$ the measure induced by the measure
$\mu$ on $\mathbf{X}$, then $\mu(t)\in\mathbf{\mathfrak{M}}$. Denote
\[
c_{t}(j,v)=c_{t}p_{t}(j,v)\]
 where $c_{t},p_{t}$ are the concentration- and the densities at
time $t$; 
\item the trajectories ($x_{i}(t),v_{i}(t),j_{i}(t))$ are independent for
different $i$ and each of them is a trajectory of some Markov process,
not necessarily time homogeneous. This process is defined bu the initial
measure $\mu(0)\in\mathfrak{M}$ and by infinitesimal transition probabilities
at time $t$, moreover the latter do not depend on coordinates, velocities
and types of other particles, but depend on $c_{t}(j,v)$ at time
$t$; 
\item the evolution of the pair $(j,v)$ for individual particle is defined
by the following Kolmogorov equation \[
\frac{\partial p_{t}(j_{1},v_{1})}{\partial t}=\]
 \[
=\sum_{j}\int(P(t;j_{1},v_{1}|j,v)p_{t}(j,v)-P(t;j,v|j_{1},v_{1})p_{t}(j_{1},v_{1}))dv\]
 where\[
P(t;j_{1},v_{1}|j,v)=\sum_{j',j'_{1}}\int2a_{r}((j_{1},v_{1}),(j_{1}',v'_{1})|(j,v),(j',v'))c_{t}(j',v')dv'dv'_{1}\]
 Note that in the reactions there can be energy conservation law,
see \cite{Ser2,Ser3}, then the integration in the last formula includes
the corresponding $\delta$-functions; 
\item coordinates of the particle $i$ change as\[
x_{i}(t)=x_{i}(0)+\int_{0}^{t}v_{i}(s)ds\]

\end{itemize}
Let the measure $\mu(0)\in\mathfrak{M}$ be given. Consider now the
sequence of finite system of particles, defined above, in an expanding
system of finite volumes $\Lambda\uparrow R^{3}$. The initial measure
for any of finite systems is the restriction of the measure $\mu(0)$
on the corresponding volume $\Lambda$. Hence, at the initial moment
we can fix a separate particle $i$ in $R^{3}$ and can consider its
trajectories $x_{i}^{(\Lambda)}(t)$ in each $\Lambda$, starting
from some. Besides this, its trajectory $x_{i}(t)\in\mathbf{X}^{0,\infty}$
is defined in $R^{3}$.

We have seen that two definitions - for finite and infinite particle
systems - are quite different. The following result shows how they
are related.

\begin{theorem}

For any $i$ and $t$ we have the convergence in probability\[
\lim_{\Lambda\uparrow R^{3}}x_{i}^{(\Lambda)}(t)=x_{i}(t)\]

\end{theorem}

Proof see in the Appendix to the paper \cite{Ser3}.

\subsection{Boltzmann equation with chemical reactions}

\subsubsection{Model and result}

Assume that at time $t=0$ in the cube $\Lambda\subset R^{d}$ there
are $N<\infty$ particles, each particle is characterized by the coordinate
$x\in\Lambda$, velocity $v\in R^{d}$ and type $q\in\left\{ 1,...,Q\right\} $.
The initial vector $(x_{i}(0),q_{i}(0),v_{i}(0):i=1,...,N)$ will
be denoted by $\omega_{0}^{\Lambda}$. Assume that the initial measure
$\mu_{\Lambda}$ on the set $\omega_{0}^{\Lambda}$ of configurations
in $\Lambda$ is defined by the probabilities $p_{N}$ that the number
of particles in $\Lambda$ equals $N$, and by the conditional densities
(for given $N$)\[
\rho_{N,\Lambda}(x_{1},v_{1},q_{1},...,x_{N},v_{N},q_{N})<C^{N}\]
 symmetric with respect to the permutation group $S_{N}$. and normalized
so that\[
\frac{1}{N!}\sum_{q_{1},...,q_{N}}\int\rho_{N,\Lambda}\prod_{k=1}^{N}dx_{k}dv_{k}=1\]
 Then the $k$-particle correlation functions are defined as\[
f_{k}(x_{1},v_{1},q_{1},...,x_{k},v_{k},q_{k})=\sum_{N=k}^{\infty}\frac{p_{N}}{(N-k)!}\sum_{q_{k+1},...,q_{N}}\int\rho_{N,\Lambda}\prod_{j=k+1}^{N}dx_{j}dv_{j}\]
 Hence, the probability that in any of small volumes $dx_{i}dv_{i},i=1,...,k$
there is a particle of type $q_{i}$, is equal to \begin{equation}
f_{k}(x_{1},v_{1},q_{1},...,x_{k},v_{k},q_{k})\prod_{i=1}^{k}dx_{i}dv_{i}\label{density}\end{equation}
 For any initial configuration $\omega_{0}^{\Lambda}$ define the
continuous time Markov process\[
\xi(t,\omega)=\xi_{N,\Lambda}(t,\omega,\omega_{0}^{\Lambda})=(x_{i}(t),v_{i}(t),q_{i}(t):i=1,2,...,N)\]
 In the defined system the randomness related to the initial (for
$t=0$) configuration is denoted by $\omega_{0}^{\Lambda}$, and the
randomness related to the random jumps is denoted by $\omega$. Heuristically
$\xi(t,\omega)$ is defined as the mixture of deterministic (piecewise
linear) dynamics (with periodic boundary conditions) for the coordinates
and random jumps for the velocities and types. These random jumps
are assumed to be binary reactions, which change velocities and types.
More exactly, the velocities and types are assumed to piecewise constant
on $[0,\infty)$; the jumps occur at random time moments\begin{equation}
0<t_{1}(\omega)<...<t_{k}(\omega)<...\label{jumpmoments}\end{equation}
 On the time intervals $t\in(t_{i},t_{i+1}]$ the particle move freely
$\frac{d^{2}x_{i}(t)}{dt^{2}}=0$, that is with constant velocities
$v_{k}(t)=v_{k}(t_{i}+0)$. Thus almost everywhere we have $v_{k}(t)=\frac{dx_{k}(t)}{dt}$.
Thus, the coordinates at time $t\in R_{+}$ are defined as\[
x_{i}(t,\omega)=x_{i}(0)+\int_{0}^{t}v_{i}(t,\omega)dt\]
 At the same time any pair of particles $i,j$ (independently of the
other pairs) on any time interval $(t,t+dt)$ change their types and
velocities with the rate (rate densities)\[
\lambda(q_{i}^{\prime},v_{i}^{\prime},q_{j}^{\prime},v_{j}^{\prime}|x_{i}(t),v_{i}(t),q_{i}(t).x_{j}(t),v_{j}(t),q_{j}(t))dt\]
 The functions $\lambda$ are assumed non-negative, bounded, smooth,
translation invariant and equal zero, if at least one of the following
conditions holds:

\begin{enumerate}
\item $|x_{i}-x_{j}|\geq2r$; 
\item for some $v^{0}>0$ either $|v_{i}^{\prime}|>v^{0}$ or $|v_{j}^{\prime}|>v^{0}$. 
\end{enumerate}
These jumps do not change coordinates, but change velocities and types\[
(q_{i},v_{i},q_{j},v_{j})=(q_{i},v_{i},q_{j},v_{j})(t)\rightarrow(q_{i}^{\prime},v_{i}^{\prime},q_{j}^{\prime},v_{j}^{\prime})=(q_{i}^{\prime},v_{i}^{\prime},q_{j}^{\prime},v_{j}^{\prime})(t+0)\]
 Denote\[
B(v,q,v^{\prime},q'|v_{1},q_{1},v_{2},q_{2})=\]
 \[
=\lambda(v,q,v^{\prime},q'|x,v_{1},q_{1},x,v_{2},q_{2})\]
 Thus, for given $N,\Lambda,\lambda,r$ we have defined the family
$\xi_{N,\Lambda}(t)$ of processes with finite number of particles.

\begin{remark}

This model allows many generalizations, for example when the movement
of particles between jumps is defined by the Hamiltonian system with
pair potential $V(x-y)$ and interaction radius $r$\[
\frac{d^{2}x_{i}(t)}{dt^{2}}=-\sum_{j:j\neq i}\frac{\partial V(x_{i}-x_{j})}{\partial x_{i}}\]

\end{remark}

\paragraph{Initial conditions}

We consider the family $\mu_{r}$ of initial distributions in a finite
volume $\Lambda$ or in $R^{3}$ with correlation functions $f_{k}^{(r)}(0;x_{1},v_{1},q_{1},...,x_{k},v_{k},q_{k})$,
parametrized by positive numbers $r$ (interaction radii). It is assumed
that this family satisfies the following conditions:

\begin{enumerate}
\item (Boltzmann-Grad scaling) For some fixed bounded non-negative function
$f(x,v,q)$ \begin{equation}
f_{1}^{(r)}(0;x,v,q)=r^{-d+1}f(x,v,q)\label{BG-scaling}\end{equation}
 This scaling says that the mean density of particles grows as $r^{-d+1}$
as $r\to0$. Note that simultaneously {}``effective'' volume $Nr^{d}$
($r$ can be considered as {}``effective'' radius of the particles),
occupied by the particles, tends to zero as $r$; 
\item (exponential decay of correlations)\[
|f_{2}^{(r)}(0;x_{1},v_{1},q_{1},x_{2},v_{2},q_{2})-f_{1}^{(r)}(0;x_{1},v_{1},q_{1})f_{1}^{(r)}(0;x_{2},v_{2},q_{2})|<\]
 \[
<C\exp(-C_{1}r^{-1}|x_{1}-x_{2}|)\]
 for all positive sufficiently small $r\to0$: 
\item (strong stability) The number of particles $n(A)$ in arbitrary volume
$A\subset\Lambda$ is uniformly bounded a.s. by $cr^{-d+1}|A|$ for
some $c>0$. It is a physically natural condition; 
\item (bounds from above) for some $C>0$\[
f_{k}^{(r)}(0;x_{1},v_{1},q_{1},...,x_{k},v_{k},q_{k})<C^{k}r^{(-d+1)k},k\geq1\]

\end{enumerate}
Let us give an example of a point field, satisfying conditions 1-4.
Consider first the Poisson field with the correlation functions\[
g_{k}(x_{1},v_{1},q_{1},...,x_{k},v_{k},q_{k})=\prod_{i=1}^{k}g_{1}(x_{i},v_{i},q_{i})\]
 for some smooth bounded functions $g_{1}$. Define $\mu_{r}$ as
the point field, obtained from the Poisson field with $k$-particle
functions $g_{k}^{(r)}=r^{(1-d)k}g_{k}$ by the transformation $F$
on the set of configurations, where the configuration $\omega_{1}=F(\omega_{0})$
is obtained from $\omega_{0}$ by deleting any particle for which
in $\omega_{0}$ there is another particle on the distance less or
equal to $2r$. Then the properties 1,3 follow from the definition.
The property 4 follows from the monotonicity\[
f_{k}^{(r)}\leq g_{k}^{(r)}\]
 The property 2 follows from standard estimates of the probability
that there exists a cluster (a sequence of particles with the distance
not more than $2r$ of subsequent particle from the previous) of the
diameter $\frac{1}{2}|x_{1}-x_{2}|$, containing at least one of the
particles $x_{1}$ or $x_{2}$.

\subsubsection{Existence of cluster dynamics}

We say that two particles $i,j$ interact at time $t$, if at this
moment a reaction occurred between them, in particular at this moment
the distance between them was not more than $2r$. Denote $s_{ij}$
the first moment of interaction of particles $i$ and $j$. Consider
the following finite random graphs $G^{\Lambda}=G^{\Lambda}(\tau)=G^{\Lambda}(\tau,\omega)$.
Their vertexes are the particles, it is convenient to enumerate them
with their initial vectors $x_{i}(0)$. Two vertexes are connected
by the edge if on the time interval $\left[0,\tau\right]$ these vertexes
interacted at least once. For fixed $\tau$ and $\Lambda$ the set
of vertexes of any connected component of the graph $G^{\Lambda}=G^{\Lambda}(\tau)$
is called a dynamical cluster (in $\Lambda$).

If at point $x$ at initial moment there is a particle then denote
$P_{k}^{\Lambda}(\tau|x)$ the conditional probability that the dynamical
cluster to which this particle belongs, consists of exactly $k$ particles.
Put\[
\rho=\max_{x}\sum_{q}\int f_{1}(x,v,q)dv\]
 Then the following exponential estimate holds, see \cite{MalJMPh}.

\begin{theorem}

Their exist constants $C,\alpha_{0}>0$ such that for any $\tau,v^{0},r$
and\begin{equation}
\rho=\alpha(\tau v^{0}r^{d-1})^{-1}\label{malost}\end{equation}
 with arbitrary $0<\alpha<\alpha_{0}$ uniformly in $k,\Lambda,x$
the following inequality holds\[
P_{k}^{\Lambda}(\tau|x)\leq(C\alpha)^{k-1}\]

\end{theorem}

Remind that the expression for the density has the following meanng:
the mean number of particles $<N>$ in the unit volume, multiplies
on maximal volume of the tube, covered by the particle (that is $2r$-neighborhood
of its trajectory), were less than some $\alpha_{0}$. Proof see in
\cite{MalJMPh}.

From this theorem it follows that with probability $1$ any particle
on the time interval $[0,\tau]$ belongs to a finite cluster depending
on the chosen particle and on the initial configuration. Moreover,
there is the thermodynamic limit which is the cluster dynamics in
$R^{d}$. Otherwise speaking, the dynamics of infinite number of particles
is obtained from infinite number of independent finite particle dynamics.

More exactly, there is the following cluster representation of the
dynamics. Consider the Markov process $\xi_{k}(t)=(x_{1}(t),v_{1}(t),q_{1}(t),...,x_{k}(t),v_{k}(t),q_{k}(t))$
on the time interval $[0,\tau]$ with $k$ particles (assuming the
absence of other particles), if initially the particles were at points
$\xi_{k}=\xi_{k}(0)=(x_{1},v_{1},q_{1},...,x_{k},v_{k},q_{k})$. Let
us denote $\gamma=\gamma(\eta|\xi)$ the trajectory of the process
$\xi_{k}(t)$. starting at time $0$ at point $\xi$ and ending at
time $\tau$ at the point $\eta=(x_{1}',v_{1}',q_{1}',...,x_{k}',v_{k}',q_{k}')$.
Denote $P^{\tau}(\gamma(\eta|\xi))$ the conditional distribution
on the set $\{\gamma(\eta|\xi)\}$ of such trajectories Let $\Gamma_{k}=\Gamma_{k}(\eta|\xi)$
be the set of all such trajectories $\gamma$, which form a $k$-cluster.

Then using (\ref{density}) we get\[
f(\tau;x,v,q)=\]
 \[
=\sum_{k=1}^{\infty}\sum_{q_{1},...,q_{k}}^{\infty}\int_{R^{dk}\times\Lambda^{k}}\prod_{i=1}^{k}dx_{i}dv_{i}\prod_{i=2}^{k}dx'_{i}dv'_{i}\int_{\Gamma_{k}}Q(\gamma)\]
 \[
dP^{\tau}(\gamma((x,v,q,x_{2}',v_{2}',q_{2}',...,x_{k}',v_{k}',q_{k}')|\xi))f_{k}(0;x_{1},v_{1},q_{1},...,x_{k},v_{k},q_{k})\]
 where $Q(\gamma)=Q(\gamma(\eta|\xi))$ is the conditional probability
that other particle in the configuration $\omega_{0}^{\Lambda}$ do
not interact with the distinguished $k$ particles.

\subsubsection{Proof of the Boltzmann equation}

If the existence of cluster dynamics holds for fixed $r$, then the
Boltzmann equation holds only in the Boltzmann-Grad limit. Fix $\tau$
and $v^{0}$ and assume that the function $f(x,v,q)$ from the condition
(\ref{BG-scaling}) is sufficiently small so that the conditions (\ref{malost})
holds.

\begin{theorem}

Then in the Boltzmann-Grad limit for any $t<\tau$ there exist the
density functions\[
\lim_{r\to0}r^{d-1}f^{(r)}(t;x,v,q)=f(t;x,v,q)\]
 which satisfy the Boltzmann equation\[
\frac{\partial f}{\partial t}(x,v,q)=-v\frac{\partial f}{\partial x}(x,v,q)+\]
 \[
+\sum_{q',q_{1},q_{2}}\int[B(v,q,v^{\prime},q'|v_{1},q_{1},v_{2},q_{2})f(x,v_{1},q_{1})f(x,v_{2},q_{2})-\]
 \[
-B(v_{1},q_{1},v_{2},q_{2}|v,q,v',q')f(x,v,q)f(x,v',q')]dv_{1}dv_{2}dv^{\prime}\]

\end{theorem}

In the zeroth approximation (that is if there is no interaction) there
are only $1$-clusters and then\begin{equation}
f(t+\delta;x,v,q)=f(t;x-v\delta,v,q)\label{0approx}\end{equation}
 Subtracting $f(t;x,v,q)$ from both sides of this inequality, dividing
by $\delta$ and passing to the limit $\delta\to0$, we get\[
\frac{\partial f}{\partial t}=-v\frac{\partial f}{\partial x}\]

In the general case the equality (\ref{0approx}) corresponds to the
event,that the particle which were at the point $x,v$ at time $t$,
did not react in the time interval $(t,t+dt)$.

In our case we can write\begin{equation}
f(t+\delta;x,v,q)=f(t;x-v\delta,v,q)-A_{1}^{(r)}+B_{1}^{(r)}+O(\delta^{2})\label{1approx}\end{equation}
 The term $A_{1}$ is obtained from the events, when at time $t$
there were a particle with parameters $x-v\delta,v,q$, which in the
sequel (in some time $s,t<s<t+\delta$) reacted with another particle,
which at time $t$ had the parameters $x_{1},v_{1},q_{1}$. In other
words\[
A_{1}^{(r)}=\sum_{q_{1},q',q_{1}^{\prime}}\int_{t}^{t+\delta}ds\int\lambda(q^{\prime},v^{\prime},q_{1}^{\prime},v_{1}^{\prime}|x(s),v(s),q(s),x_{1}(s),v_{1}(s),q_{1}(s))\]
 \[
f_{2}^{(r)}(t;x-v\delta,v,q;x_{1},v_{1},q_{1})dx_{1}dv_{1}dv'dv_{1}^{\prime}\]
 Note that as it follows from the cluster property, the decay of correlations
is conserved for any time moment in the interval $0<t\leq\tau$. Then
\[
A_{1}^{(r)}\to\delta\sum_{q_{1},q',q_{1}^{\prime}}\int B(q^{\prime},v^{\prime},q_{1}^{\prime},v_{1}^{\prime}|v,q,v_{1},q_{1})f_{1}(q,v)f_{1}(q_{1},v_{1})dv^{\prime}dv_{1}^{\prime}dv_{1}\]
 The term $B_{1}^{(r)}$ appears from the events, when at time $t$
there are two particles with parameters $x_{1},v_{1},q_{1},x_{2},v_{2},q_{2}$,
which react at time $s,t<s<t+\delta$ so that one of emerging particles
has parameters$x,v,q$. Thus,\[
B_{1}^{(r)}=\sum_{q_{1},q_{2},q'}\int_{t}^{t+\delta}ds\int\lambda(v,q,q^{\prime},v^{\prime}|x_{1},v_{1},q_{1},x_{2},v_{2},q_{2})\]
 \[
f_{2}^{(r)}(x_{1},v_{1},q_{1},x_{2},v_{2},q_{2})dx_{1}dv_{1}dx_{2}dv_{2}dv'\]
 Similarly we have\[
\lim_{r\to0}B_{1}^{(r)}=\delta\sum_{q_{1},q_{2},q'}\int B(v,q,q^{\prime},v^{\prime}|v_{1},q_{1},v_{2},q_{2})f_{1}(q_{1},v_{1})f_{1}(q_{2},v_{2})dv_{1}dv_{2}dv'\]

From the cluster representation of the dynamics it easily follows
that the remaining clusters do not contribute to the Boltzmann equation,
as they the order $O(\delta^{2})$.

\begin{remark}

Other models and other techniques of proving the Boltzmann equation
see in \cite{Lanf,Spohn,CaMaPul,CaprPul,CaprPul-2,CaPulWag,CerIllPul,CerPet}.

\end{remark}

\subsection{Simplest models with transport on the lattice}

Let at each point $x$ of the lattice $Z^{d}$ there can be $n_{v}(x)$
particles of type $v=1,2,...,V$. In each point the Markov process
$\xi_{x}=(n_{1}(x),...,n_{V}(x))$ is given, that is a chemical reaction
system as defined in section 2. These processes are independent and
have the same distribution. We add some terms to the generator of
this process, which correspond to independent simple continuous time
random walk for each of the particles, homogeneous in time and space.
The parameters $\lambda_{e,v}$, that is the jump rates, where $e$
runs $2d$ unit vectors along the axes,- can depend only of the type
$v$. Assume now that the drift vectors\[
m_{v}=\sum_{e}e\lambda_{e,v}\neq0\]
 for all $v$. We will use the scaling

\[
x=\frac{X}{\epsilon},t=\frac{\tau}{\epsilon},\lambda_{r}(n)\to\epsilon\lambda_{r}(n)\]
 where $X\in R^{d},\tau\in R$ are macro-variables. In the definition
(\ref{intensity}) we put $M=1$, so the number of particles in any
point has the order $O(1)$. This scaling says in particular, that
for finite macro-time $\tau$ at a given point there can be $O(\tau)$
reactions.

\begin{theorem}

Let at the initial moment $t=0$ the initial Poisson distribution\[
\prod_{x}\prod_{v}\frac{(b_{v,x})^{n_{v}(x)}}{n_{v}(x)!}exp(-b_{v,x})\]
 of the particles on the lattice so that $b_{v,x}=c_{v}(0,\epsilon x)$
for some smooth bounded functions $c_{v}(0,X),X\in R^{d}$. Then as
$\epsilon\to0$ for any functions $x(\epsilon):R_{+}\to Z^{d}$ such
that $\epsilon x(\epsilon)\to X$, there exist the limits of the concentrations
\[
c_{v}(\tau,X)=\lim_{\epsilon\to0}<n_{v}(\frac{\tau}{\epsilon},x(\epsilon))>\]
 which satisfy the equations\begin{equation}
\frac{\partial c_{v}}{\partial\tau}=-m_{v}\frac{\partial c_{v}}{\partial X}+F_{v}(c_{1},...,c_{V})\label{eulerR}\end{equation}
 where the functions $F_{v}$ are the same as in the right-hand side
of the equations (\ref{kinetic_equ}).

\end{theorem}

Shortly, the ideas of the proof are as follows. Firstly, it is well-known
and easy to prove, that under no reaction condition the independent
particle satisfies the equation (\ref{eulerR}) without the last term.
The reactions go much slower than the transport and for finite macro-time
their number at each point is $O(1)$. Thus, at the intervals between
reactions the distributions of different particle types at each point
tend to become independent and have Poisson distribution with some
parameters $c_{v,x}$, due to the fast mixing by random walks. Thus
in the limit $\epsilon\to0$ for any integer $k>0$\[
<n_{v,x}(n_{v,x}-1)...(n_{v,x}-k+1)>_{Poisson}=c_{v,x}^{k}\]
 The estimates for the convergence of limit transitions are based
on one or another variant of cluster expansions. See more details
of the proofs in \cite{Spohn}, pp. 308-313, 315-316 and in the references
therein.

If all drifts $m_{v}=0$, then one needs another (diffusion) scaling\[
x=\frac{y}{\epsilon},t=\frac{\tau}{\epsilon^{2}},\lambda_{r}(n)\to\epsilon^{2}\lambda_{r}(n)\]
 which corresponds to the difference of the scales of reaction times
and temperature movement. Here we get reaction-diffusion equations,
if we assume the jump rates in each direction $\lambda_{v}=\frac{1}{2}$,\[
\frac{\partial c_{v}}{\partial\tau}=\frac{1}{2}\Delta c_{v}+F_{v}(c_{1},...,c_{V})\]
 The ideas of the proof are similar to the previous theorem.

There is an interesting case when the coordinates of the drift are
nonzero in one direction (current) and zero for perpendicular directions
(diffusion). Let us consider, for example, the two-dimensional lattice
$Z^{2}=\{(x,y)\}$ with several particle types, where $m_{v,x}=0,m_{v,y}\neq0$
for all $v$. Then in the scaling\[
x=\frac{X}{\epsilon},y=\frac{Y}{\epsilon^{2}},t=\frac{\tau}{\epsilon^{2}},\lambda_{r}(n)\to\epsilon^{2}\lambda_{r}(n)\]
 the limiting equations are\[
\frac{\partial c_{v}}{\partial\tau}=-m_{v,X}\frac{\partial c_{v}}{\partial Y}+\frac{1}{2}\frac{\partial^{2}c_{v}}{\partial X^{2}}+F_{v}(c_{1},...,c_{V})\]

All kinetic equations, described above, may have several invariant
measures if the corresponding equations\[
F_{v}(c_{1},...,c_{V})=0\]
 have several fixed points.

\section{Chemical reaction as a process}

\subsection{Time reversibility in physics}

A map (function, operator) $f$ from the set $A$ onto the set $B$
is called invertible, if it is one-to-one, that is if there exists
mapping $g=f^{-1}$ of the set $B$ onto the set $A$ such that $gf$
is identical on $A$, then $fg$ is identical $B$. Then $f$ and
$g$ are called mutually inverse. If $A=B$, then $f$ is called automorphism.

Automorphism $U$ generates the automorphism group $U^{t}$, where
$t\in Z$, and it is often possible to embed it into some continuous
automorphism group with $t\in R$. The automorphism group of the set
$A$ is called time invertible with respect to the automorphism $W$
of the set $A$, if\begin{equation}
W^{-1}U^{t}WU^{t}=1\label{physObrat}\end{equation}
 or $U^{-t}=W^{-1}U^{t}W$.

In physics the notion of time invertibility (reversibility) is related
to concrete automorphisms (normally involutions) $W$. So; in the
classical Newtonian physics of $n$ particles one considers the automorphism
groups of the manifold\[
\Lambda^{n}\times R^{nd}=\{(\overline{x},\overline{p})=(x_{1},...,x_{n},p_{1},...,p_{n})\},x_{i}\in\Lambda\subset R^{d},p_{i}\in R^{d},\]
 and $W$ is taken equal \[
W(\overline{x},\overline{p})=(\overline{x},-\overline{p})\]
 In non-relativistic quantum mechanics $A$ is a complex Hilbert space,
automorphism group is a unitary group $U^{t}=\exp(itH)$ with the
generator $H$, and $W$ is an anti-linear map. For example, in some
representation\[
W\phi=\overline{\phi}\]
 If $HW=WH$, then (\ref{physObrat}) holds.

In relativistic quantum theory invertibility may take place for one
operators $W$, but nor for the other, as for parity violation.

Such physical invertibility, under certain conditions, implies reversibility
of corresponding stochastic systems. Example is the reversibility
of the transfer matrix in the euclidean approach to quantum field
theory, the invertibility of the scattering matrix and reversibility
of Markov processes, obtained in a weak interaction limit.

The scattering matrix $S:\mathcal{F}\rightarrow\mathcal{F}$ is the
unitary operator in the Fock space $\mathcal{F}=\mathcal{F}(\mathcal{H})$,
where $\mathcal{H}$ is the direct sum of all one-particle subspaces
$\mathcal{H}{}_{q}$ in $\mathcal{F}(\mathcal{H})$. One-particle
subspace corresponds to the type $q$ particle. Let $e_{q,k}$ be
some orthonormal basis in $\mathcal{H}_{q}$, where $k$ corresponds
to momentum. Let $\gamma=\{(q_{1},k_{1}),...,(q_{n},k_{n})\}$ and
consider the basis \[
e_{\gamma}=e_{q_{1}k_{1}}\otimes...\otimes e_{q_{n}k_{n}},n=0,1,...,\]
 in $\mathcal{F}$. Denote the squares of the matrix elements of $S$
in this basis by\[
\left|S(\gamma\to\gamma')\right|^{2}=w(\gamma\to\gamma')\]
 Then from the norm invariance it follows that\begin{equation}
\sum_{\gamma'}w(\gamma\to\gamma')=\sum_{\gamma'}w(\gamma'\to\gamma)=1\label{unita}\end{equation}
 that is the matrix $W=(w(\gamma\to\gamma'))$ is doubly stochastic.

Note that the condition (\ref{unita}) corresponds to the unitarity
condition introduced in section 2. Intuitively, to get stochastic
dynamics from unitary scattering matrix one could do as follows: introduce
random waiting times in the states $\gamma$, then matrix $W$ will
play role of the matrix of conditional probabilities (at the end of
waiting time) of jumps from $\gamma$ to $\gamma'$. There is one
more difficulty - $\gamma$ depend on index $k$, which correspond
to momenta, or to velocities. It is desirable that the process, obtained
by restriction to the set of types $\{ q\}$ would be Markov. And
moreover, that the matrix $W^{0}$, which is obtained by restricting
the matrix $W$ to the set of types, also be doubly stochastic. This
would mean that there is restoration of classical chemical kinetics,
which is normally defined without taking into account velocities.
In \cite{Ser4} it is shown that under some conditions this really
happens.

Under certain limiting transitions unitary quantum dynamics becomes
stochastic, for example in the weak interaction limit, see for example
\cite{Hepp}, moreover one has reversible Markov process.

\subsection{Classical Hamiltonian scattering}

The first question is whether any bound state can be obtained dynamically
by colliding its components. It is not always possible. It is simpler
to see from the classical scattering theory, already for the simplest
example of scattering of one particle on the center, see \cite{ReedSimon3}.
The main restriction is of course the energy conservation law, that
is of the sum of internal and kinetic energies. We do not know whether
it is the unique restriction.

\paragraph{Scattering of one particle on external potential }

Let us consider one-dimensional problem of scattering of one classical
particle on smooth external potential $V(x),x\in R$, equal zero outside
some bounded interval. Let the particle move from $-\infty$ with
speed $v_{0}$. It is known that the formation of the bound state,
that is the capture of the particle by the potential, is possible
only for the set velocities of measure zero. This follows from the
time invertibility of the dynamics. In fact, three cases are possible:

1) the particle changes the direction of movement. This can be only
in the point where $\frac{dV}{dx}\neq0$. Then the particle passes
the same path in the opposite direction;

2) the particle stops. This can occur only in the points $\frac{dV}{dx}=0$.
The number of such values of $v_{0}$ is finite.

3) the particle passes to $+\infty$. Then it does not stop and does
not change direction.

There similar results, see \cite{ReedSimon3}, for the one particle
scattering on external potential for dimensions $d\geq2$ .

If however one may introduce the possibility of fast or momentary
energy dissipation, for example when the energy is transferred to
the third particle or to some internal degrees of freedom of the two
particle cluster, then the formation of the bound state becomes possible.
The particle can fall into the potential energy well.

\paragraph{Two-particle and three-particle scattering }

Two-particle scattering can be reduced to the previous case, and for
three particles there are examples from celestial mechanics, see \cite{Alekseev},
which show what possibilities are possible.

\subsection{Reaction rates from local theory}

The following simple model shows how one can obtain reaction rates
from local models.

Consider a free system of $N_{i},i=1,2,$ balls of radii $r_{i}$
in the volume $M$, which move not seeing each other with Maxwell
velocity distribution, that is in equilibrium. Denote $c_{i}=\frac{N_{i}}{M}$
the fractions of such particles. It is easy to prove that the number
of binary reactions {}``collisions'' of balls of different types
on a large time interval $[0,t]$ is asymptotically equal to (if $N_{i},M\to\infty,$)\[
R(N,M,t)\sim tN_{1}N_{2}\frac{4\pi(r_{1}+r_{2})^{3}}{3M}\]
 Thus, the mean number of collisions for u nit time and volume is
equal to\[
\frac{R}{tM}=\frac{4\pi(r_{1}+r_{2})^{3}}{3}c_{1}c_{2}\]
 The part of energetic collisions, that is such that the sum of kinetic
energies is not less than some number $T_{cr}$, is approximately\[
a_{r}\exp(-\beta T_{cr})c_{1}c_{2}\]
 for some constant $a_{r}$, corresponding to a given reaction. This
is the contents of the Arrhenius law.

Various mathematical problems, related to the transfer of chemical
energy to kinetic energy see in \cite{Ser2,Ser3,Ser4}.

\subsection{Dynamics with non-momentary reactions}

Let us consider the infinite system of point particles in $R^{d}$,
where each particle is defined by the coordinate $x$, velocity $v$
and type $j=1,...,J$. The particles interact via pair potentials
$U_{jj'}(x-x')$. It is assumed that the interaction radii are $R_{jj'}$finite.
Define the graph with vertexes in the particle coordinates, connecting
two particles of types $j,j'$ with an edge, if the distance between
them does not exceed some $R_{jj'}$. To the vertex also we prescribe
velocity and type of the corresponding particle. Consider connected
finite marked graphs $G$ with vertexes marked in this way. There
are continuum of such graphs, and we introduce the equivalence classes
of such graphs. Two graphs $G_{1}$ and $G_{2}$ are called equivalent,
if at least one of the following three conditions holds:

\begin{enumerate}
\item $G_{1}$ and $G_{2}$ are isomorphic as marked graphs, that is they
are isomorphic as graphs and the marks of the corresponding vertexes
coincide; 
\item $G_{1}$ and $G_{2}$ are obtained from one another by shifting all
coordinates with the same vector; 
\item $G_{1}$ and $G_{2}$ are obtained from one another by Hamiltonian
dynamics of the corresponding particle system with potentials $U_{jj'}(x-x')$
for finite time. 
\end{enumerate}
The equivalence class will be called cluster or metastable particle.
Metastable particle is called stable, or a bound state, if it exists
for infinite time under the absence of other particle.

Infinite system of particles in $R^{d}$ at any time moment can be
partitioned on (maximal) connected components. Each component corresponds
to a cluster.

Assume now that there exists Gibbs equilibrium state for the given
particle system (in particular, that the potential is stable). Then
under some conditions, for example for small density of particles,
one can prove that at any time moment all clusters are finite with
probability 1. Moreover, some extended clusters (called above the
dynamical clusters) stay finite on some on some time interval, see
\cite{MalJMPh}.

However, other problems, for example about the existence times of
clusters are open. For example, it seems that (stable) bound states
are absent with probability 1, except of course one-particle clusters.
In fact, let we have for example only two types of particles $1,2$
and the potential $V_{12}$ is such that the unique possible bound
state consists of two particles, where one of the particles rotates
around another with the circle orbit. Some relation between velocity
of the rotation and the distance (radius of the circle orbit) between
particles. But, as the distributions of coordinates and velocities
are independent, this is possible only with zero probability.

From other side, there exist infinite number of metastable clusters,
where these relations hold approximately, and clusters live sufficiently
long and disappears either by itself or after collision with other
cluster. Interesting problem is to estimate the life time of a cluster
as depending of the parameters of the Gibbs distribution.

However, there are many other open questions concerning connection
of the notion of particle in finite and infinite systems.

In finite particle systems, classical and quantum, one first introduces
elementary particles. In the classical case as the point particles,
in quantum case as creation-annihilation operators. The Hamiltonian
can have also bound states. In classical case these are the orbits
of some type, in quantum - eigenfunctions of the discrete spectrum.
In some cases one can show that they completely determine possible
asymptotic states of the system. In principle, bound states of any
number of particles are possible.

If we consider infinite system, then there exists a certain theory
only for equilibrium dynamics, that is the dynamics while the system
is in the equilibrium (Gibbs) state. Here also elementary particles,
called bare or unrenormalized particles, which however do not belong,
contrary to the finite systems, to the discrete spectrum after the
thermodynamic limit. New discrete spectrum can appear (more exactly,
one particle states). The main hypothesis, proved in some cases, is
that the pair (Hilbert space, Hamiltonian), corresponding to given
infinite interacting system of particles, is unitary equivalent to
some similar pair for a system of non-interacting particles, which
are called then quasi-particles. These quasi-particles can be close
to bare particles for example if the interaction is small. In physics
a quasi-particle is often imagined as a particle surrounded by a cloud
of other bare particles. This could give a bridge between quasi-particles
in infinite systems and metastable states of finite systems, however
there are no exact formulations and proofs.


\begin{thebibliography}{10}
\bibitem{ReedSimon3}M. Reed, B. Simon. Methods of modern mathematical
physics. Volume 3, 1979. Academic Press.

\bibitem{Murray}J. Murray. Lectures on nonlinear differential equations
models in biology. 1981. Oxford.

\bibitem{McAAr}H. McAdams, A. Arkin. Stochastic mechanisms in gene
expression. Proc. Natl. Acad. Sci. 1997, 94, 814.

\bibitem{ArRoAd}Arkin A., Ross J., McAdams H. Stochastic kinetic
analysis of developmental pathway bifurcation in phage $\lambda$-infected
Escherichia coli cells. Genetics, 1998, 149, 1633.

\bibitem{GadLeeOth}C. Gadgil, Chang-Hyeong Lee, H. Othmer. A stochastic
analysis of first-order reaction networks. Preprint, 2003.

\bibitem{Leo}M. A. Leontovich. Main equations of kinetic theory of
gases from the random processes point of view. J. of Experim. and
Theor. Physics, 1935, v. 5, No. 3-4, 211-231.

\bibitem{McQua}D. McQuarrie. Stochastic approach to chemical kinetics.
J. Appl. Prob., 1967, v. 4, 413-478.

\bibitem{Kal}Kalinkin A.V. Markov branching processes with interaction.
Uspehi Mat. Nauk, 2002, v. 57, no. 2, pp. 23-84.

\bibitem{LanLif}E. Lifschits, I. Pitaevskij. Course of theoretical
physics, v. 10, Moscow.

\bibitem{BerezKlesov}I. V. Berezin; A. A. Klesov. Practical course
of chemical and fermentative kinetics. 1978, Moscow State University.

\bibitem{Kelly}F. Kelly. Reversibility and stochastic networks. 1979,
Wiley, New York.

\bibitem{Whittle}P. Whittle. Systems in stochastic equilibrium. John
Wiley. 1986.

\bibitem{Nelson}R. Nelson. The Mathematics of Product Form Queuing
Networks. ACM Computing Service, v. 25, No. 3, 1993, 339-369.

\bibitem{EthKurtz}S. Ethier, Th. Kurtz. Markov processes characterization
and convergence. 1986. J. Wiley.

\bibitem{BKPR}K. Ball, Th. Kurtz, L. Popovic, G. Rempala. Asymptotic
Analysis of multiscale approximations to reaction networks. 2005.
Preprint. arXiv: math.PR/0508015.

\bibitem{Alekseev}V. M. Alekseev. Lectures on Celestial Mechanics.
1999.

\bibitem{Ser1}V. Malyshev, S. Pirogov, A. Rybko. Random walks and
chemical networks. Moscow J. Math., v. 2, 2004.

\bibitem{Ser2}G. Fayolle, V. Malyshev, S. Pirogov. Stochastic chemical
kinetics with energy parameters. In {}``Trends in Mathematics'',
v. 3, 2004.

\bibitem{Ser3}V. Malyshev. Microscopic Models for Chemical Thermodynamics.
J. of Stat. Physics, 2005, v. 119, No. 5/6, 997-1026.

\bibitem{Ser4}V. Malyshev. V. Malyshev. Fixed Points for Stochastic
Open Chemical Systems. MPRF, 2005, v. 11, No. 2, 337-354.

\bibitem{MalJMPh}V. Malyshev. Dynamical clusters of infinite particle
dynamics. Journal Math. Phys., 2005, v. 46, No. 7.

\bibitem{Spohn}H. Spohn. Large scale dynamics of interacting particles.
1991. Springer.

\bibitem{Spohn2}H. Spohn. In {}``Studies in Statistical Mechanics'',
vol. 10, 1983.

\bibitem{DeMasiPres}A. DeMasi, E. Presutti. Lectures on the Collective
Behavior of particle systems. CARR Reports on Mathematical Physics,
1989, No. 5.

\bibitem{ArnThe}L. Arnold, M. Theodosopulu. Deterministic limit of
the stochastic model of chemical reactions with diffusion. Adv. Appl.
Prob., 1980, v. 12, 367-379.

\bibitem{BraLeb}M. Bramson, J. Lebowitz. Spatial structure in low
dimensions for diffusion limited two-particle reactions. Ann. Appl.
Prob., 2001.

\bibitem{Hepp}K. Hepp. Results and problems in Irreversible statistical
mechanics of open systems. Lecture Notes in Mathematics, v. 39, pp.
138-150. Springer. 1975.

\bibitem{Volk}M. Volkenstein. Physics and Biology. 1980. Moscow.

\bibitem{CerIllPul}C. Cercignani, R. Illner, M. Pulvirenti. The mathematical
theory of dilute gases. Springer. 1994.

\bibitem{CaprPul}S. Caprino, M. Pulvirenti. A cluster expansion approach
to a one-dimensional Boltzmann equation in a stationary state. Comm.
Math. Phys., 1995, 166, 603-631.

\bibitem{CaprPul-2}S. Caprino, M. Pulvirenti. The Boltzmann-Grad
limit for a one-dimensional Boltzmann equation: a validity result.
Comm. Math. Phys., 1995, 166, 603-631.

\bibitem{CaPulWag}S. Caprino, M. Pulvirenti, W. Wagner. Stationary
particle systems approximating stationary solutions to the Boltzmann
equation. SIAM J. Math. Anal., 1998, v. 29, No. 4, 913-934.

\bibitem{Lanf}O. Lanford. Time evolution of large classical systems.
Lecture Notes in Physics, 1975, v. 38, Springer, pp. 1-111.

\bibitem{CaMaPul}E. Caglioti, C. Marchioro, M. Pulvirenti. Non-equilibrium
Dynamics of Three-Dimensional Infinite Particle System. Comm. Math.
Phys., 2000, v. 215, No. 1, 25-43.

\bibitem{CerPet}C. Cercignani, D. Petrina, V. Gerasimenko. Many-Particle
Dynamics and Kinetic Equations. 1997. Kluwer. 
\end{thebibliography}
\end{document}